\documentclass[single]{cambridge6Atight}
\usepackage{aas_macros}
\usepackage{natbib}
\usepackage{rotating}
\usepackage{floatpag}
\rotfloatpagestyle{empty}
\usepackage{graphicx}
\usepackage{multind}
\usepackage{amssymb}
\usepackage{times}

\newcommand{\teff}{$T_{\mathrm{eff}}$}

\newcommand{\numax}{$\nu_{\mathrm{max}}$}
\newcommand{\dnu}{$\Delta\nu$}
\newcommand{\dP}{$\Delta P$}

\newcommand{\wire}{\textit{WIRE}}
\newcommand{\kepler}{\textit{Kepler}}

\newcommand{\keplermission}{\textit{Kepler Mission}}

\begin{document}

\chapter{Asteroseismology of red giant stars}
\begin{center}
{\normalsize Rafael A. Garc\'\i a$^1$ \& Dennis Stello$^2$\\
\vspace{0.3cm}
$^1$Laboratoire AIM, CEA/DSM -- CNRS - Univ. Paris Diderot -- IRFU/SAp, Centre de Saclay, 91191 Gif-sur-Yvette Cedex, France\\
$^2$Sydney Institute for Astronomy (SIfA), School of Physics, University of Sydney, NSW 2006, Australia
}
\end{center}

\section{Introduction}

If appropriately excited, a star will oscillate like a giant spherical
instrument. In most stars, including the Sun, surface convection provides
the excitation mechanism \citep{1977ApJ...212..243G}.  With turbulent
velocities reaching speeds comparable to the local sound speed near the
surface of the star,
the vigorous convective motions can excite standing acoustic waves.  These
are known as pressure or p modes because the restoring force arises from
the pressure gradient.     
The broad frequency spectrum of this excitation mechanism gives rise to many
oscillation modes, both radial and non-radial, excited
simultaneously. These stochastically excited and intrinsically damped
oscillations were first detected in the Sun \citep{1962ApJ...135..474L},
and hence are commonly known as solar-like oscillations.  

Oscillation modes can be characterised by the
number of nodes, $n$, in the radial direction, called the radial order, and a
non-radial part described by spherical harmonics, each with an angular
degree, $l$, which equals the number of nodal lines on the surface, and an
azimuthal order, $m$, which is the number of those nodal lines crossing the
equator of the star. Except for the Sun, we can genrally not resolve these
oscillations on the surfaces of cool stars and so the surface 
displacements of modes with high angular degree ($l\gtrsim4$) cannot be
observed because regions of opposite phase tend to cancel out.   

\begin{figure}[hrtb!]
\includegraphics[scale=0.55]{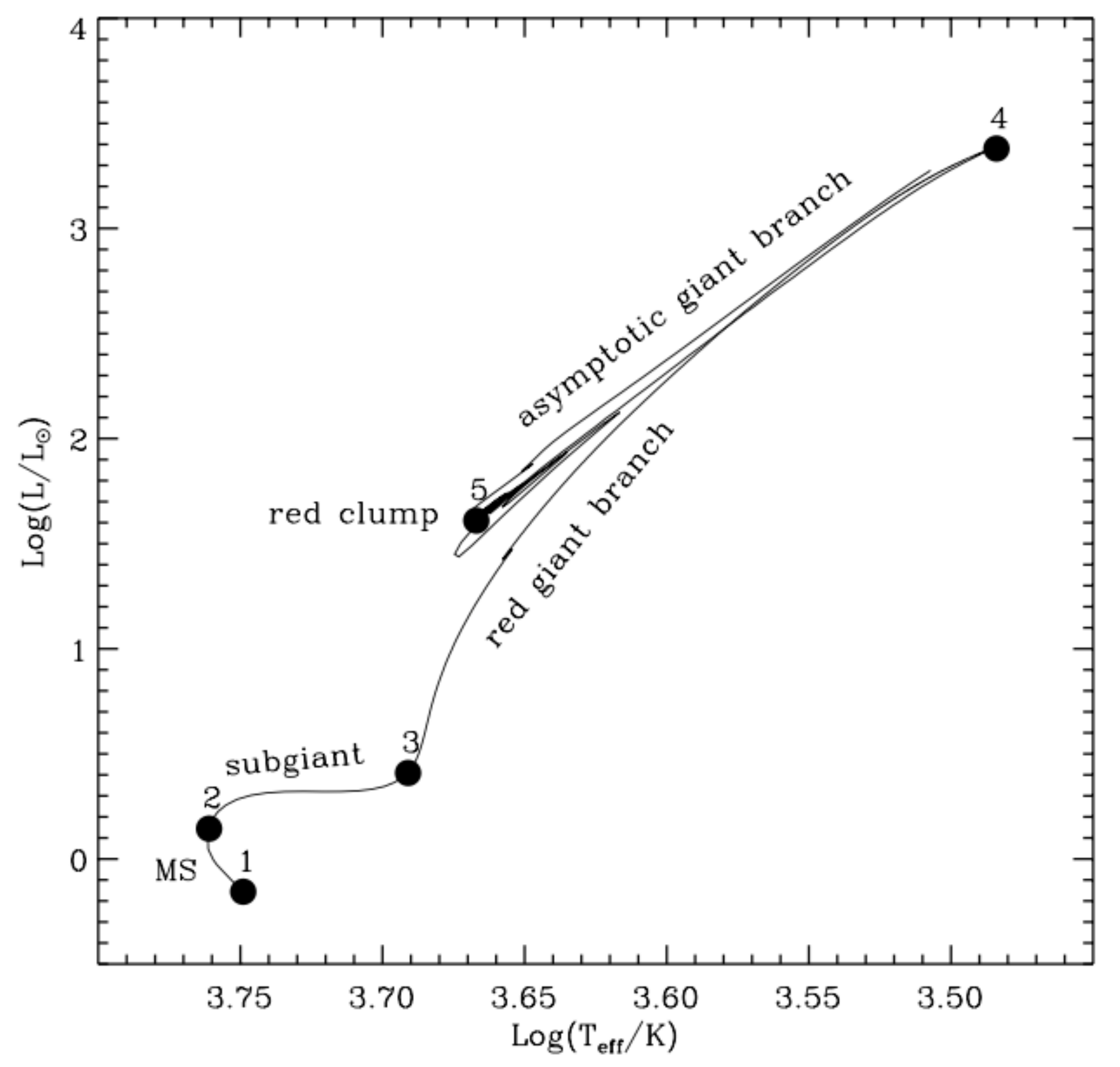}
\caption[HR diagram of a one solar mass star]
{HR diagram of a stellar model for a one solar mass star of solar
  composition, showing luminosity in solar units versus effective (surface)
  temperature. The star starts its life burning hydrogen at the 
  zero-age-main-sequence (point 1). Each following evolution phase is indicated as follows:
  1-2: main sequence (hydrogen core burning); 2-3: subgiant (hydrogen shell
  burning); 3-4: red giant branch (hydrogen shell burning); 4: red giant
  branch tip (helium ignition); 5: red clump (helium core burning); 5 and up: asymptotic giant
  branch (helium shell burning).}
\label{hrd}
\end{figure}

When stars grow old and the supply of hydrogen fuel is exhausted in the
core, their envelopes expand and cool: they become subgiants and eventually
red giants. 
This transition is shown in the so-called HR-diagram in Fig~\ref{hrd}, which
indicates the main stages of evolution of a one solar mass star from the
'main sequence' where the Sun is currently located (between points 1-2)
through to the red giant phase discussed in this chapter (beyond point 3).
Like the Sun, red giants have convective outer envelopes but the much
longer convective time scales drives oscillation modes at much lower
frequencies. The expansion and contraction of different parts of the
stellar surface when a star oscillates gives rise to variations in temperature -- and
hence also luminosity -- across the stellar surface. For the Sun,
the surface moves with about walking speed and has an oscillation amplitude of
the order of 100 m.  The corresponding brightness variations are
only of a few parts per million (ppm), equivalent to temperature
fluctuations of about 0.1 K.  In most red giants, those brightness variations
have increased to several tens, even hundreds of ppm, which is however,
still relatively small compared to the twinkling of stars we see at night
caused by the Earth's atmosphere. From the ground, the most successful
method for detecting the oscillations has therefore been using time-series
spectroscopy to detect the surface velocity changes, through the Doppler shift of
the light, as the stellar surface moves in and out.  This method is not
impacted by the twinkling of stars but current limitations on the ultra
high-precision spectrographs required for these observations generally
restrict investigations to one target at a time.  Hence, to do large scale
investigations one needs to use photometry to measure the
brightness of many stars all at the same time, which requires space-based observations.  

Before asteroseismology entered the space age, a range of ground-based
efforts were carried out using either spectroscopy or
photometry. Measurements of solar-like oscillations were pursued since the
early 1990s \citep[e.g.][~and references herein]{1993AJ....106.2441G}, but it was nearly a
decade later before we saw the first firm detection in a red giant
\citep{2002A&A...394L...5F}. Despite an increasing number
of oscillating red giants detected from ground \citep[e.g.][]{2004ESASP.559..113B,2006A&A...448..689D} and early space missions
\citep{2000ApJ...532L.133B,2003ApJ...591L.151R,2007A&A...468.1033B,2008AJ....136..566G}, at the
time, it was still not clear whether these stars showed both radial and
non-radial modes \citep{2004SoPh..220..137C}, and hence how strong the mode
damping was \citep{2006A&A...448..709S}.  It was therefore not known how information
rich and useful red giants would be as seismic targets.  In an ambitious attempt
to detect oscillations in a large group of red giants in the open cluster
M67 using 10 telescopes simultaneously for over one month, only marginal
detections were achieved, which concluded almost two decades of
ground-based photometric attempts \citep{2007MNRAS.377..584S}.   

\section{Red giants: the new frontier in asteroseismology}\label{frontier}

\begin{figure}[hrtb!]
\includegraphics[scale=0.55]{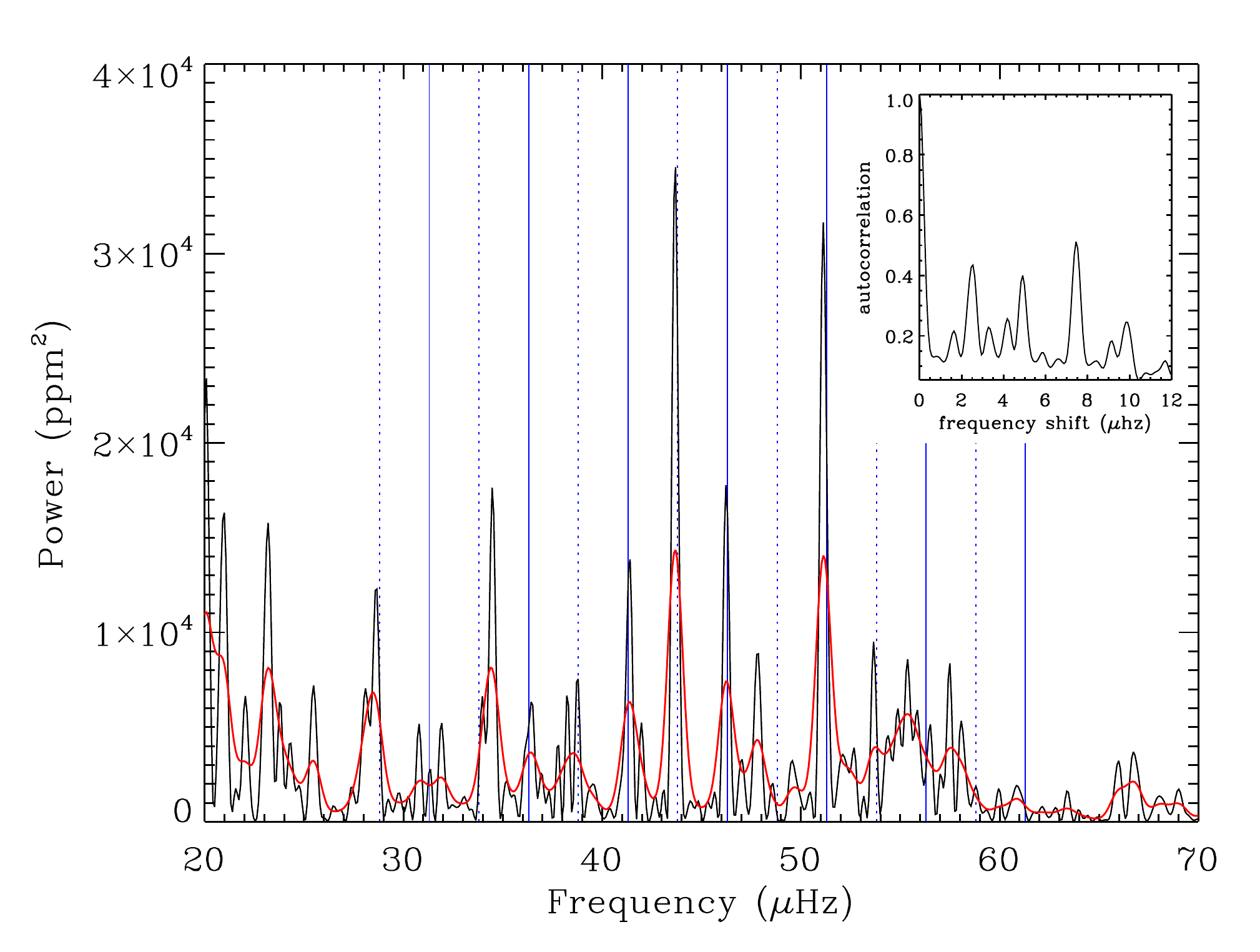}
\caption[Power spectrum of the red giant $\beta\,$Vol]
{Power spectrum of the red giant $\beta$Vol observed by the star tracker on
  the \wire\ spacecraft (smoothed
  version in red). Equally spaced vertical solid and dotted lines indicate
  radial and dipole modes, respectively.  The inset shows the
  autocorrelation of the spectrum, showing peaks at multiples of half of the
  large frequency spacing: \dnu/2 (Stellar Pulsation and Evolution 2007, Vancouver).}
\label{wire}
\end{figure}

It was data from space missions that transformed the field and demonstrated
that red giants show rich spectra of oscillation frequencies. The existence
of non-radial modes  (see Fig.~\ref{wire}) were indicated in a few bright stars
observed by the star tracker camera on the \wire\ spacecraft
\citep{2008ApJ...674L..53S}. 
But it was a sample of hundreds of stars
observed by CoRoT that unambiguously 
demonstrated that red giants exhibit both radial and non-radial modes
\citep{2009Natur.459..398D}. Shortly after this came the initial results
from the \keplermission, which provided more precise data for
even larger numbers of stars
\citep{2010PASP..122..131G,2010ApJ...713L.176B}, confirming they pulsate
with both radial and non-radial modes, whose frequency patterns resembled
that of the roughly equally-spaced acoustic modes in the Sun, but at lower
frequencies. However, in a few percent of stars the dipole
modes showed unexplained low amplitudes
\citep{2012A&A...537A..30M,2014A&A...563A..84G}. Over a period of only 2-3
years, the number of red giants with detected 
oscillations went from a handful (pre-CoRoT) into the thousands
\citep{2010A&A...517A..22M,2010A&A...522A...1K,2011MNRAS.414.2594H,2011ApJ...743..143H,2012EPJWC..1905012M,2013ApJ...765L..41S}.
These long high-precision time series for large numbers of stars opened up a
completely new way of exploring red giants using asteroseismology. \\ 

\begin{figure}[hrtb!]
\includegraphics[scale=0.45]{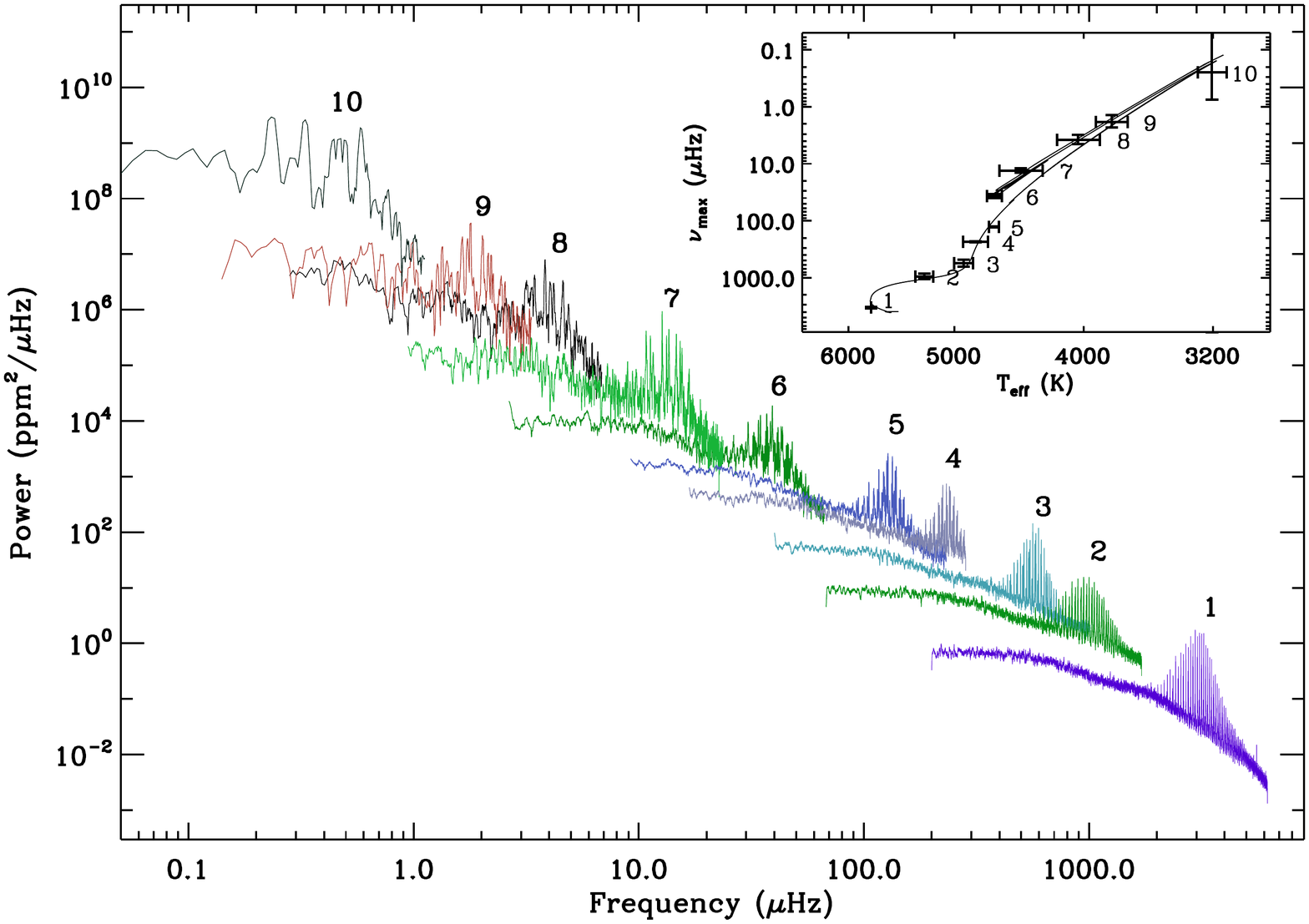}
\caption[Seismic HR diagram and sequence of stellar oscillation spectra.]
{\label{PSDs} Sequence of stellar oscillation spectra starting from a main
  sequence star, the Sun ($\#1$), to the tip of the red giant branch. Star
  $\#6$ is a red clump star. Temperatures are from the
  \emph{Kepler} Input Catalog \citep{2011AJ....142..112B} and from
  \citet{2014ApJS..211....2H}. The inset represents a seismic HR
  diagram. The solid line is the evolutionary track of a one solar-mass
  stellar model computed
  using MESA \citep{2011ApJS..192....3P,2013ApJS..208....4P} in which the
  position of each star is represented. The 1-$\sigma$ error bars in $\nu_{\rm{max}}$
  are multiplied by a factor of 10.} 
\end{figure}

The large frequency separation computed from the equally-spaced overtone modes provide
information on the density of the star. Through scaling relations
anchored at the solar values, this information combined with the
frequency of maximum power, $\nu_{\rm{max}}$, and the effective temperature allows estimation
of mass and radius for stars pulsating from minutes to hundreds of days
(See Fig.~\ref{PSDs}).  These relations have been extensively tested for
both main-sequence and red giant stars with additional constraints such as
parallax measurements or individual frequencies 
\citep{2008ApJ...674L..53S,2011arXiv1107.1723B,2012MNRAS.419.2077M,2012ApJ...749..152M}
and by using stellar models \citep{2009MNRAS.400L..80S,2011ApJ...743..161W,2013EPJWC..4303004M,2013ASPC..479...61B}.
Recent independent methods have verified the scaling relations at the 4--5\% level
\citep{2011ApJ...743..143H,2013MNRAS.433.1262W,2012ApJ...757...99S}, although
uncertainties remain about additional systematics for giants
at certain stages of their evolution, which need further investigation
\citep{2012MNRAS.419.2077M,2013EPJWC..4303004M}. 
Masses and radii can therefore be inferred for a wide range of evolutionary
stages from the main sequence through to the red giant branch, the clump, and even
asymptotic giant branch stars \citep{2012ApJ...757..190C}.  

Among the most luminous red giants observed by \kepler, it was showed that
also they pulsate in both radial and non-radial modes
\citep{2013A&A...559A.137M}, concluding a decade-long debate about the possible 
presence of non-radial modes in these highly evolved stars.  Stello et
al. 2014[REF IN BIB] further demonstrated that the frequency pattern of these stars was
markedly different from that of lesser evolved stars with dipole modes
dominating the power frequency spectrum and being shifted relative to the
radial modes. 


\section{Mixed modes: a window into the inner radiative core of red giants}

\begin{figure}[hrtb!]
\includegraphics[scale=0.24]{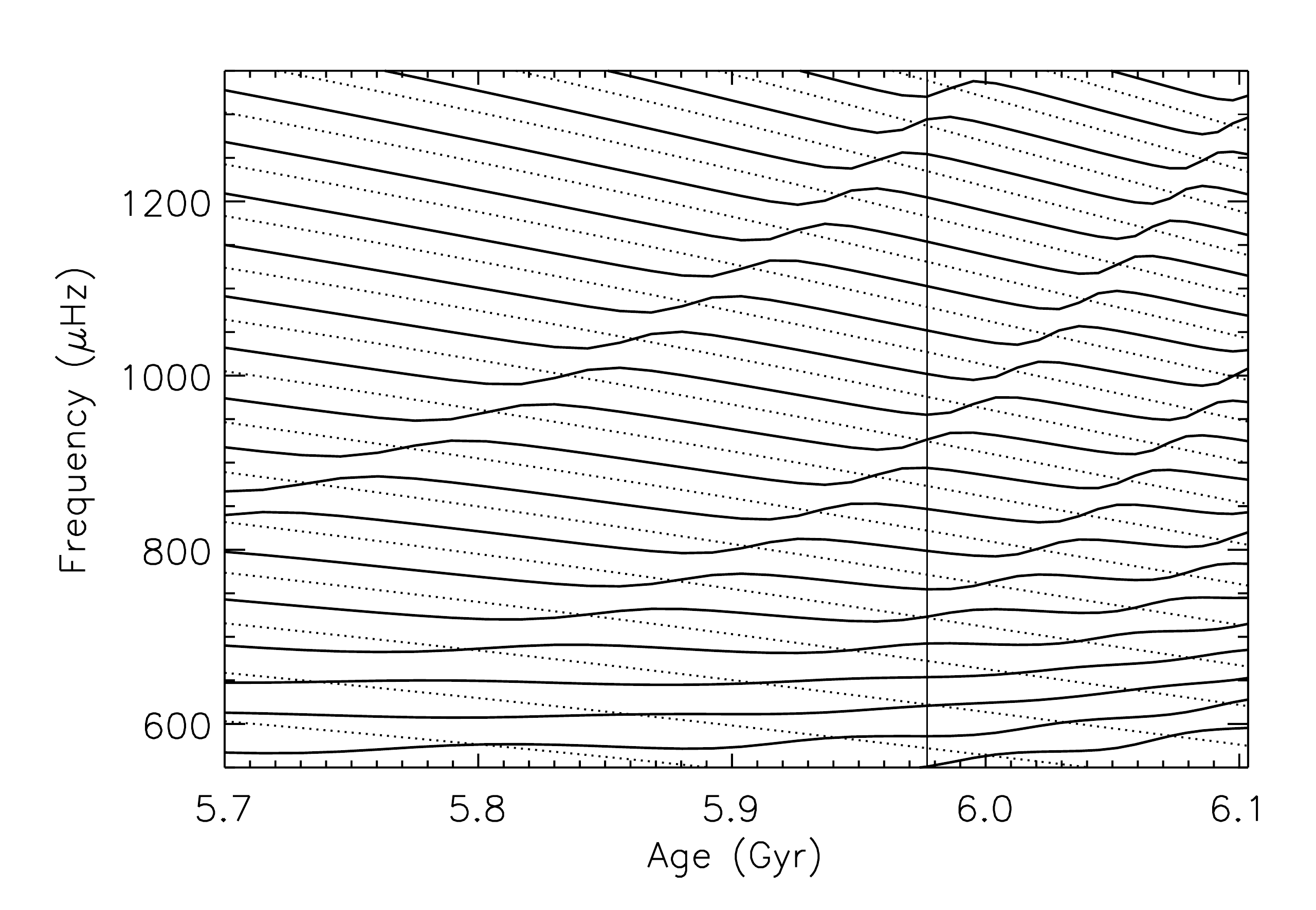}
\includegraphics[scale=0.24]{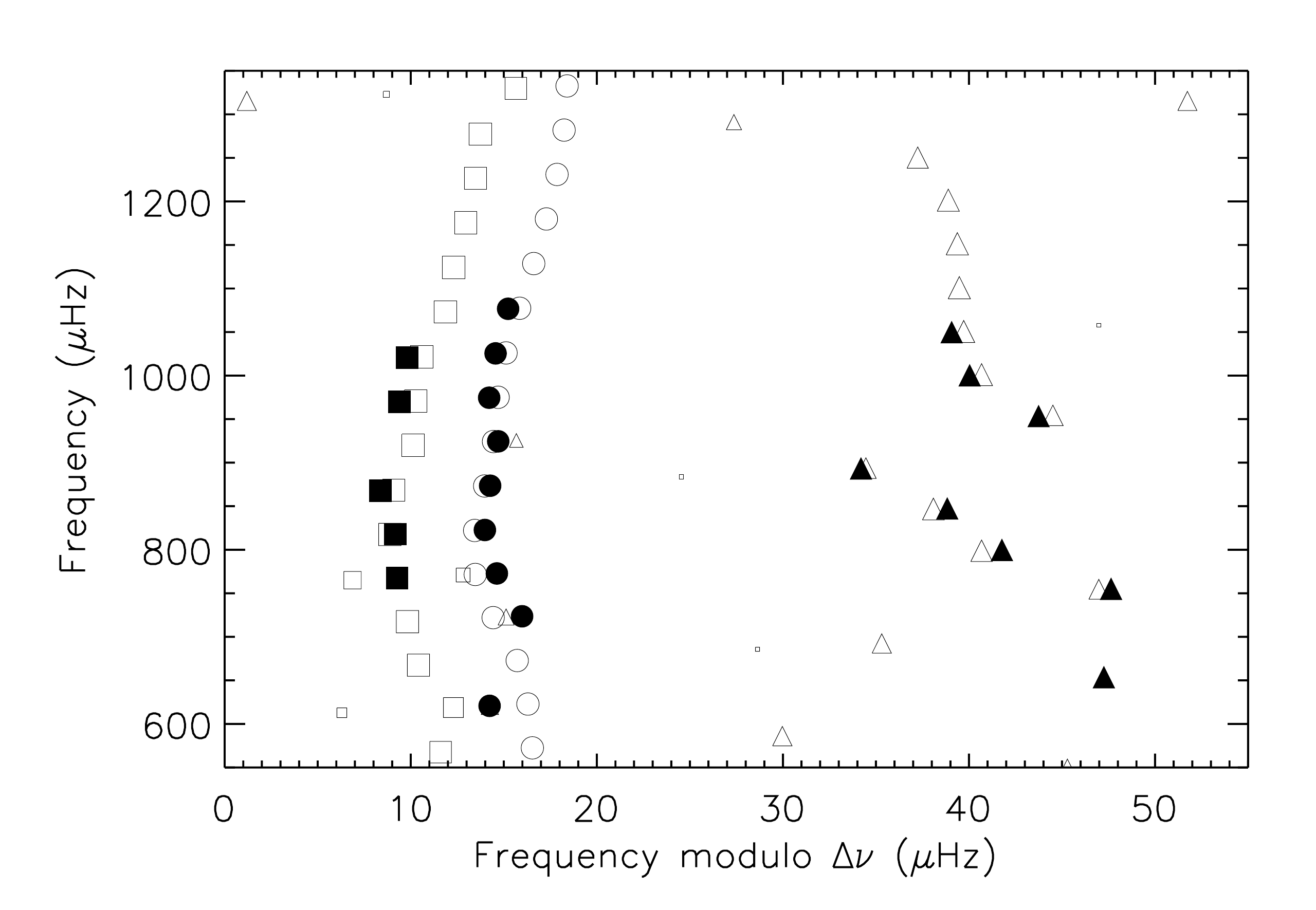}
\caption[Example of mixed modes in a subgiant star]
{
Left: Temporal evolution of the $l=0$ (dotted) and $l=1$ (solid)
oscillation frequencies based on KIC~11026764
\citep{2010ApJ...723.1583M}. The vertical solid line indicates the location
of the model whose frequencies matches the observed. Right: \'Echelle
diagram using a frequency separation of $\Delta\nu$ = 50.5 $\mu$Hz. Open
symbols are the frequencies from the best fitting model. Solid symbols are the observed frequencies from \citet{2010ApJ...723.1583M}. Circles are used for modes with $l = 0$, triangles for $l = 1$, and squares for $l = 2$ modes (Figure courtesy of Dr. G. Do\u{g}an). }
\label{bump}
\end{figure}

One of the key reasons red giants have been such a success story reaching
even beyond the
field of asteroseismology in recent years is the presence of so called
mixed modes, predicted by \citet{2009A&A...506...57D} and later observed by \citet{2010ApJ...713L.176B}. 
In the following, we will explain how these modes occur. The stellar
interior of cool stars essentially consists of two cavities, an outer
envelope where standing acoustic waves (p modes) predominantly reside, and
a radiative core where standing gravity waves (g modes) propagate. For a star like the Sun, p-mode frequencies
are much higher than those of the g modes -- 3000 $\mu$Hz or 5 minutes
compared to 100 $\mu$Hz or 2.8 hours, respectively. They therefore do not
`sense' the presence of each other.  When a star evolves to become a red
giant the core contraction and the envelope expansion results in the p-mode
frequencies becoming lower while the g-mode frequencies become higher and
eventually resulting in overlapping p- and g-mode frequencies.  This causes the two types of waves to couple.
When a p mode couples with a g mode and form a mixed mode its frequency is
shifted -- the mode is bumped from its original position -- as the two modes
undergo a so-called `avoided crossing'. The regular frequency spacing
between consecutive modes of the same degree and different order is
broken. The first mixed modes occur during the subgiant phase where the  
associated mode bumping can be observed. Examples include stars like $\eta$
Boo \citep{1995AJ....109.1313K,2005A&A...434.1085C}, $\beta\,$Hyi
\citep{2007ApJ...663.1315B,2011A&A...527A.145B}, the CoRoT subgiant
HD~49385 \citep{2010A&A...515A..87D}, and many \emph{Kepler} targets
\citep[e.g.][]{2011ApJ...733...95M,2012ApJ...749..152M,2011A&A...534A...6C}.
An example, based on the \emph{Kepler} target KIC~11026764
\citep{2010ApJ...723.1583M}, is shown in Fig.~\ref{bump}. The left-hand
panel shows the temporal evolution of radial and dipole modes of a series
of models. Each instance where the dipole p mode (solid line) moves up in
time, we have an avoided crossing with an underlying g mode whose presence
can be traced by the series of mode bumpings of ever increasing frequency. The model
that best matches the observations is marked with a vertical line. The
corresponding \'echelle diagram is showed in the right-hand panel, which is expected
to show near vertical ridges for modes of the same spherical degree $l$,
meaning almost equidistant frequencies following the typical pattern
of high-order low-degree acoustic modes described by the asymptotic
relation (REF  TASSOUL).  
Bumped dipole modes are clearly seen around 900 $\mu$Hz, revealing the
presence of a g mode at that frequency. 

When stars reach the red giant phase, the frequency spectrum of the
g modes become denser than that of the p-modes. Hence, for each p
mode there are several g modes with the same angular degree
and similar frequencies, each coupling to the p mode leading to several
mixed modes with a p mode character in the envelope and a g mode character
in the core \citep{1975PASJ...27..237O,1977A&A....58...41A,1991A&A...248L..11D,2011Sci...332..205B}. 
Because of the coupling, mixed modes carry valuable information about the
inner radiative interior directly to the surface, and due to their
p-mode character, their amplitudes are high enough to be detected at the surface. 
This is in contrast to main sequence stars like the Sun where g modes reach
the stellar surface with tiny amplitudes because these modes are evanescent in the
convective envelope, making them difficult to detect \citep{2007Sci...316.1591G}. 

The closer in frequency a g mode is to a p mode the more p-mode dominated
the resulting mixed modes will be.  A series of g modes coupling to the
same p mode therefore generates mixed modes with different degree of p- and g-mode
character, each probing different regions of
the stellar interior, allowing a stratified `view' of the stellar interior. 


\section{Period spacings: a way to distinguish different stellar evolution phases}

Classical observations of the surface properties of red
giants do not allow us to disentangle stars ascending the red giant branch -- only burning
hydrogen in a shell -- from stars in later evolution stages also burning helium in
their cores such as the red clump stars (Fig~\ref{HRD}).

By measuring the period spacing of the dipole mixed modes,
\citet{2011Natur.471..608B} realized that red giants fall into two well defined
groups, those with period spacings around 50 seconds (hydrogen-shell-burning stars) and those with period spacings ranging between 100 to 300
seconds (stars also burning helium in the core). The reason for this divide
arise because g modes senses the difference in the core as it becomes
convective and expands due to the onset of the helium burning. 
\citet{2012A&A...541A..51K} demonstrated that the constant term of a linear
fit to the radial modes versus their radial order also enabled them to 
characterise this divide.  

\begin{figure}[!htb]
\includegraphics[scale=0.35]{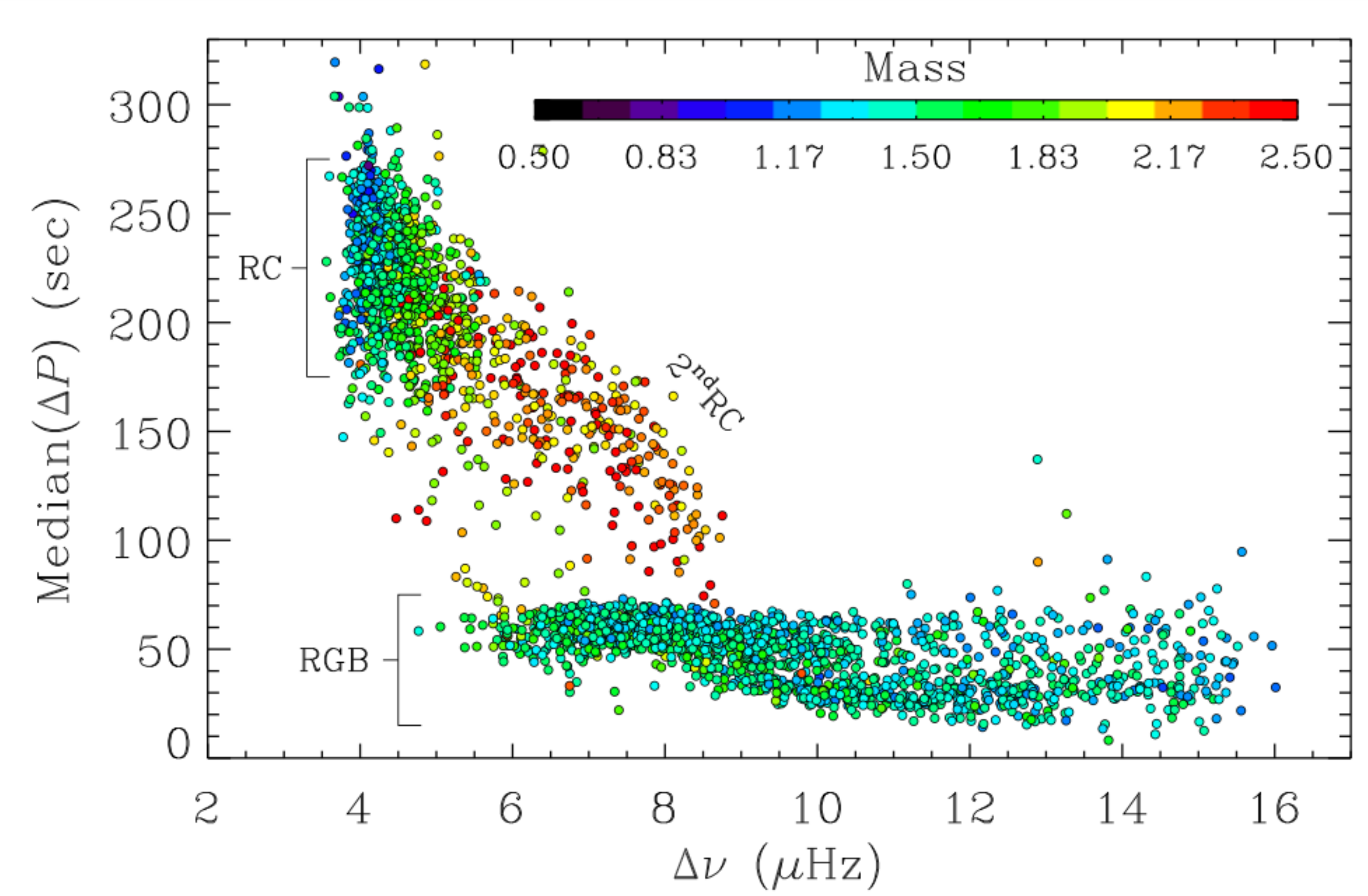}
\caption[Mixed-mode period spacing versus large frequency separation]
{Median period spacing of mixed modes for each star analyzed by \citet{2013ApJ...765L..41S} versus its large frequency separation. Red giant branch (RGB), red clump (RC), and secondary clump (2$^{\rm{nd}}$RC) stars are indicated. The mass of the stars obtained from the scaling relations are color coded.}
\label{DP}
\end{figure}

This characterisation can now be done automatically for very large samples of
stars. In Fig.~\ref{DP} we show the period spacing, \dP, against the large
frequency separation of several thousand red giant stars observed by
\kepler\ \citep{2013ApJ...765L..41S}. Red giant branch stars have period
spacings below $\sim$ 80 s. Red clump
stars are mostly concentrated above 180 s, being the less massive stars
clustering on the left-hand side of the diagram (lower $\Delta\nu$), when
they reach this evolutionary stage. The more massive helium-core-burning stars in the clump
form the high-\dnu\ low-\dP\ tail of the helium core burning stars (marked 2$^{\rm{nd}}$RC).  

\begin{figure}[hrtb!]
\includegraphics[scale=0.48]{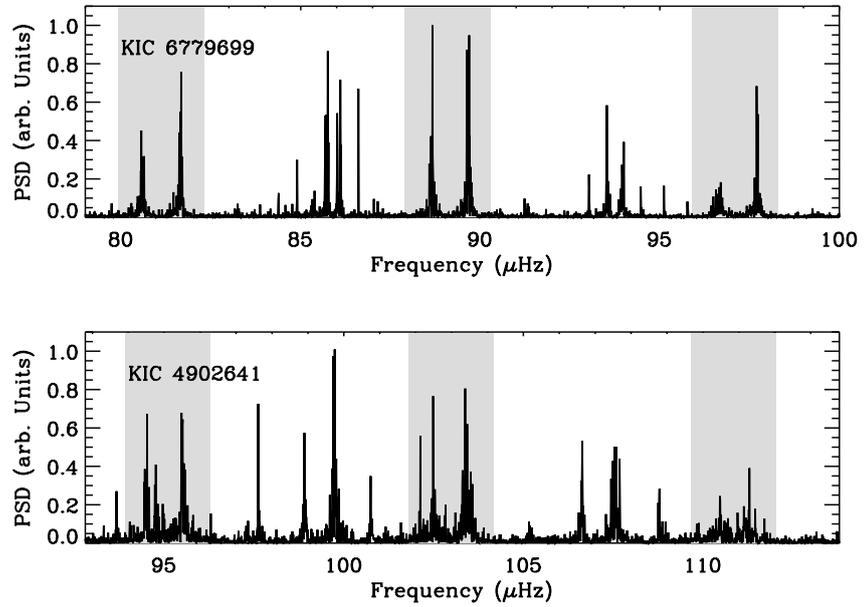}
\caption[Example of the power spectrum of 2 RG stars: a RGB and a RC]
{Normalized power spectral density (PSD) of two red giant stars, one on the red giant branch (top) and the other in the red clump (bottom).  Shaded regions correspond to the $l=0, 2$ modes, while the unshaded regions are dominated by dipole mixed modes.}
\label{FigDP}
\end{figure}

 In Fig.~\ref{FigDP} we show the power spectrum around the frequency of
 maximum power, $\nu_{\rm{max}}$, of two red giants observed during $\sim$4 years by \emph{Kepler}, one on the
 red giant branch and the other in the red clump \citep[for more details see][]{2012A&A...541A..51K}. The red clump star shows a relatively large spacing between consecutive dipole modes (large $\Delta P$) while the red giant branch star has its dipole modes bunched closely together (small $\Delta P$).

It was further noted by \citet{2013ApJ...766..118M} that \dP\ to some degree
follows the helium core mass for stars that has just ignited helium in their
cores.  This potentially gives a way to measure the amount of mixing that took
place in these stars during their earlier evolution stages. 

\section{Differential rotation inside red giants}

Mixed modes not only allow us to extract the structure of the inner core of
red giants, they also probe the distribution of internal angular momentum during the evolution along the red giant branch. Information about the angular momentum distribution is inaccessible to direct observations, but it can be inferred from the effect of rotation on oscillation frequencies. 
An extensive overview on surface and internal rotation of stars can be
found in chapter 5.6. Here we will give only a brief explanation of the
most important discoveries related to red giants.

\begin{figure}[hrtb!]
\includegraphics[scale=0.29]{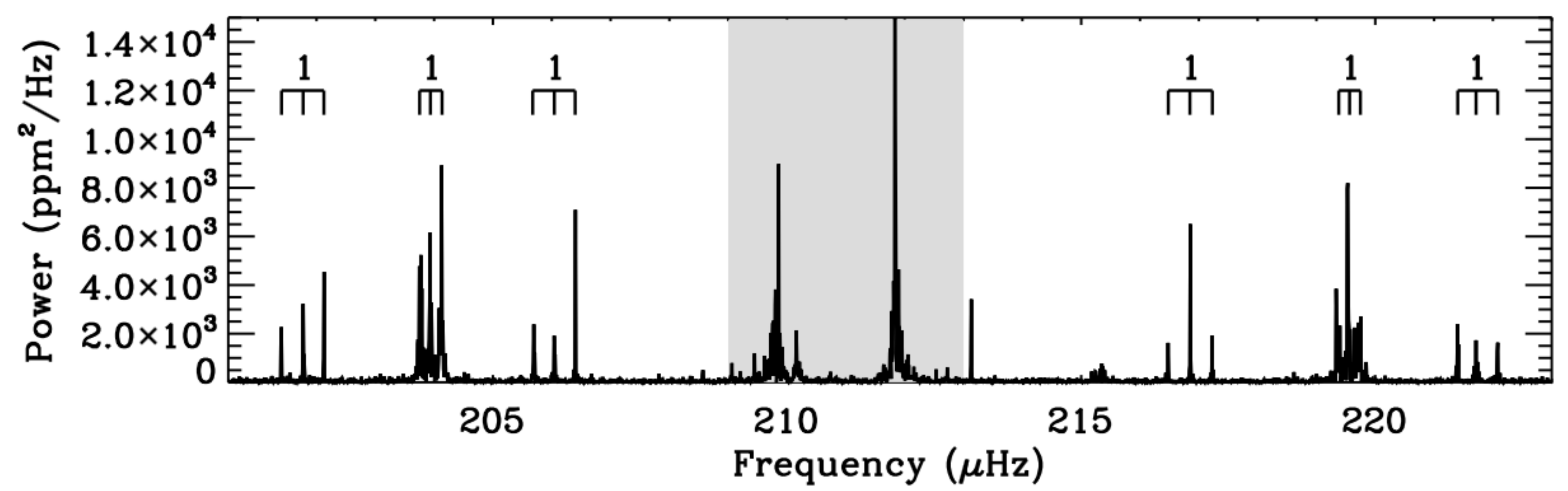}
\caption[Example of rotational splitter modes in a RG star]
{Power density spectrum of KIC~5356201 observed during nearly 4 years by \emph{Kepler}. Rotationally split $l=1$ mixed modes are marked. The shaded regions covers the $l=0,2$ modes.}
\label{Rota}
\end{figure}

\citet{2012Natur.481...55B} detected non-rigid rotation in the interior of three {\it Kepler} red giants by exploiting the rotational frequency splitting of mixed modes. An example of the power spectrum of one of these stars, KIC~5356201, can be seen in Fig.\ref{Rota}.
\citet{2012ApJ...756...19D} performed an inversion of the
radial rotation profile in an early red giant, KIC~7341231, and recently \citet{2014arXiv1401.3096D} analysed
the rotation of six subgiants and red giants within a range of
metallicities and masses. 
The latter study demonstrated that
the stellar cores spun up during the subgiant phase, while the envelopes
spun down. For two of the stars, the radial rotation profile showed a
discontinuity located at the depth of the hydrogen-burning shell, which
roughly corresponds to the boundary between the contracting and the expanding layers.

The ensemble analysis of around 300 red giants (see the next section for more details on ensemble studies), performed by \citet{2012A&A...548A..10M} showed the evolution of the rotation rate of the radiative cores from the subgiant phase to the red clump and the 2$^{\rm{nd}}$RC. It demonstrated that the rotational splitting of dipole mixed modes is mostly sensitive to the core rotation \citep{2013A&A...549A..75G}.
In the earlier stages of the red giant branch, the core rotation shows a slight decrease as the stars evolve. This seems to be in contradiction with core contraction when only local conservation of angular momentum is assumed \citep[e.g.][]{2012A&A...544L...4E,2013A&A...555A..54C,2013A&A...549A..74M}. 
Later, when stars reach the red clump, the cores have spun down even
more. Part of this effect is related to the expansion of the helium burning
core \citep{1971PASP...83..697I,2000ApJ...540..489S}.  Nevertheless, this
spin down of the core during the red giant branch and to the red clump seems to be in
agreement with the slow rotation rate found in low-mass white dwarfs
\citep{1999ApJ...516..349K}, which are essentially the end product of red
giant cores after the envelope has been stripped away during the late
stages of red giants.

\section{Ensemble Asteroseismology and Stellar populations}\label{ensemble}
With
more than 16,000 red giants observed by \emph{Kepler} and about the same
number observed by CoRoT, the space age of asteroseismology has opened up for the statistical analysis of large samples of stars usually called `ensemble asteroseismology'. Helped by these large numbers, we can identify new and common features in asteroseismic diagrams revealing key properties of stellar evolution, while providing ideal tracers of Galactic stellar evolution history to relative large distances \citep{2009A&A...503L..21M}. 

Using the scaling relations for \dnu\ and \numax\ \citep{1995A&A...293...87K} on ensembles can provide valuable results on the mass and radius
distributions of large cohorts of stars.
Aided by stellar modeling, such ensemble analysis can also provide
estimates of the age distributions. When combining radius with the apparent stellar brightness we can further estimate distances out to several kiloparsecs (1 parsec $\sim$ 3 light years). In comparison, precise distances currently available from direct parallax measurements \citep{2007ASSL..350.....V} reach only a few hundred parces. Hence, for red giants or any other
solar-like oscillating star it is possible to write:
\begin{equation}
\log d =1+2.5 \log \frac{T_{\rm{eff}}}{T_{\rm{eff},\odot}} + \log \frac{\nu_{\rm{max}}}{\nu_{\rm{max},\odot}}-2 \log \frac{\Delta\nu}{\Delta\nu_\odot} + 0.2(m_{\rm{bol}}-M_{\rm{bol},\odot})  ,
\end{equation}


\noindent where the distance $d$ is expressed in parsecs, $m_{\rm{bol}}$ is
the apparent stellar bolometric magnitude, and $M_{\rm{bol},\odot}$ is the
absolute solar bolometric magnitude, \teff\ is the effective (surface)
temperature, and $\odot$ denotes the solar values \citep{2013MNRAS.429..423M}. Therefore,
it is possible to use red giants to map and date the Galactic disc in the
regions probed by space-borne missions such as CoRoT and \kepler, and in
the future TESS \citep{2010AAS...21545006R} and PLATO
\citep{2013arXiv1310.0696R}. The results from CoRoT and \emph{Kepler}
compared to models of synthetic populations of stars already revealed
significant differences in the stellar mass distributions. One
interpretation of these results explains this as due to the vertical
gradient in the distribution of stellar masses (hence ages) in the galactic
disk \citep{2013MNRAS.429..423M}, as well as on the star-formation rate,
and the initial mass function of stars
\citep{2009A&A...503L..21M,2011Sci...332..213C}. 

Another important contributor to galactic evolution is the rate of
mass-loss when the stars evolve as red giants. There is currently no adequate physical theory on the mass loss occurring at late stages of the red giant branch that can properly explain observations. Ensemble
asteroseismology can provide important information on this crucial phase of
stellar evolution by comparing the mass distribution of red clump stars with stars
on the red giant branch (see next section). Using CoRoT and \emph{Kepler} observations, it has
been shown that there is a population of low-mass red clump stars which is absent
on the red giant branch \citep{2011A&A...532A..86M}. Assuming the scaling relations used
to infer those masses are not introducing any systematic effects of such
sort, this result could imply that these now low-mass red clump stars lost a significant
fraction of their mass while ascending on the red giant branch. Therefore, after the
helium flash, these stars appear in the red clump with a much lower mass than
the one they had during the red giant branch.

\section{Asteroseismology of open-cluster red giants}
Open clusters have long served as universal calibrators in astronomy
because stars in a cluster are considered a homogeneous sample. They are
supposed to be formed from the same cloud of gas, which means they have
roughly the same age, common initial composition, and are at the same
distance. In stellar physics, these assumptions reduce the number of usual
unknowns when matching stellar models to observations, which in combination with precise
asteroseismic measurements promises more rigorous investigation of stellar
models. 

The scaling relations for \dnu\ and \numax\ mentioned in sections
~\ref{frontier} and ~\ref{ensemble}
allows us to assign membership of stars to a particular cluster.  The
relations depend on stellar mass, surface temperature, and luminosity,
but only the latter will be markedly different from star to star among red
giants in a cluster.  Hence, \dnu\ and \numax\ are expected to form
tight relations with apparent brigness, which for cluster members is a
good proxy for luminosity, and outliers are therefore likely non-members
due to them not sharing the cluster distance. Because 
oscillations only depend on the stellar properties, this technique is
independent of the distance to the cluster and not very sensitive to the
interstellar absorption and reddening \citep{2011ApJ...739...13S}.  
Another important success of asteroseismology of open clusters was the
determination of the distances to NGC~6791 and NGC~6819 based on the
seismic studies of red giant branch stars using grid-based modeling
\citep{2011ApJ...729L..10B}. Finally, \citet{2012MNRAS.419.2077M} estimated
the integrated red giant branch mass loss in NGC~6791 by comparing the average masses of
stars in the red clump and on the red giant branch.


\bibliography{./BIBLIO}

\begin{thebibliography}{78}
\expandafter\ifx\csname natexlab\endcsname\relax\def\natexlab#1{#1}\fi
\expandafter\ifx\csname selectlanguage\endcsname\relax
  \def\selectlanguage#1{\relax}\fi

\bibitem[\protect\citename{{Aizenman} {et~al.}, }1977]{1977A&A....58...41A}
{Aizenman}, M., {Smeyers}, P., and {Weigert}, A. 1977.
\newblock {Avoided Crossing of Modes of Non-radial Stellar Oscillations}.
\newblock {\em \aap}, {\bf 58}(June), 41.

\bibitem[\protect\citename{{Barban} {et~al.}, }2004]{2004ESASP.559..113B}
{Barban}, C., {De Ridder}, J., {Mazumdar}, A., {et~al.} 2004 (Oct.).
\newblock {Detection of Solar-Like Oscillations in Two Red Giant Stars}.
\newblock {Page  113 of:} {Danesy}, D. (ed), {\em SOHO 14 Helio- and
  Asteroseismology: Towards a Golden Future}.
\newblock ESA Special Publication, vol. 559.

\bibitem[\protect\citename{{Barban} {et~al.}, }2007]{2007A&A...468.1033B}
{Barban}, C., {Matthews}, J.~M., {De Ridder}, J., {et~al.} 2007.
\newblock {Detection of solar-like oscillations in the red giant star
  {$\epsilon$} Ophiuchi by MOST spacebased photometry}.
\newblock {\em \aap}, {\bf 468}(June), 1033--1038.

\bibitem[\protect\citename{{Basu} {et~al.}, }2011]{2011ApJ...729L..10B}
{Basu}, S., {Grundahl}, F., {Stello}, D., {et~al.} 2011.
\newblock {Sounding Open Clusters: Asteroseismic Constraints from Kepler on the
  Properties of NGC6791 and NGC6819}.
\newblock {\em \apjl}, {\bf 729}(Mar.), L10.

\bibitem[\protect\citename{{Beck} {et~al.}, }2011]{2011Sci...332..205B}
{Beck}, P.~G., {Bedding}, T.~R., {Mosser}, B., {et~al.} 2011.
\newblock {Kepler Detected Gravity-Mode Period Spacings in a Red Giant Star}.
\newblock {\em Science}, {\bf 332}(Apr.), 205.

\bibitem[\protect\citename{{Beck} {et~al.}, }2012]{2012Natur.481...55B}
{Beck}, P.~G., {Montalban}, J., {Kallinger}, T., {et~al.} 2012.
\newblock {Fast core rotation in red-giant stars as revealed by
  gravity-dominated mixed modes}.
\newblock {\em \nat}, {\bf 481}(Jan.), 55--57.

\bibitem[\protect\citename{{Bedding}, }2011]{2011arXiv1107.1723B}
{Bedding}, T.~R. 2011.
\newblock {Solar-like Oscillations: An Observational Perspective}.
\newblock {\em arXiv:1107.1723}, July.

\bibitem[\protect\citename{{Bedding} {et~al.}, }2007]{2007ApJ...663.1315B}
{Bedding}, T.~R., {Kjeldsen}, H., {Arentoft}, T., {et~al.} 2007.
\newblock {Solar-like Oscillations in the G2 Subgiant {$\beta$} Hydri from
  Dual-Site Observations}.
\newblock {\em \apj}, {\bf 663}(July), 1315--1324.

\bibitem[\protect\citename{{Bedding} {et~al.}, }2010]{2010ApJ...713L.176B}
{Bedding}, T.~R., {Huber}, D., {Stello}, D., {et~al.} 2010.
\newblock {Solar-like Oscillations in Low-luminosity Red Giants: First Results
  from Kepler}.
\newblock {\em \apjl}, {\bf 713}(Apr.), L176--L181.

\bibitem[\protect\citename{{Bedding} {et~al.}, }2011]{2011Natur.471..608B}
{Bedding}, T.~R., {Mosser}, B., {Huber}, D., {et~al.} 2011.
\newblock {Gravity modes as a way to distinguish between hydrogen- and
  helium-burning red giant stars}.
\newblock {\em \nat}, {\bf 471}(Mar.), 608--611.

\bibitem[\protect\citename{{Belkacem} {et~al.}, }2013]{2013ASPC..479...61B}
{Belkacem}, K., {Samadi}, R., {Mosser}, B., {Goupil}, M.-J., and {Ludwig},
  H.-G. 2013 (Dec.).
\newblock {On the Seismic Scaling Relations Delta nu - rho and nu max - nu c }.
\newblock {Page ~61 of:} {Shibahashi}, H., and {Lynas-Gray}, A.~E. (eds), {\em
  Astronomical Society of the Pacific Conference Series}.
\newblock Astronomical Society of the Pacific Conference Series, vol. 479.

\bibitem[\protect\citename{{Belloche} {et~al.}, }2011]{2011A&A...527A.145B}
{Belloche}, A., {Schuller}, F., {Parise}, B., {et~al.} 2011.
\newblock {The end of star formation in Chamaeleon I?. A LABOCA census of
  starless and protostellar cores}.
\newblock {\em \aap}, {\bf 527}(Mar.), A145.

\bibitem[\protect\citename{{Brown} {et~al.}, }2011]{2011AJ....142..112B}
{Brown}, T.~M., {Latham}, D.~W., {Everett}, M.~E., and {Esquerdo}, G.~A. 2011.
\newblock {Kepler Input Catalog: Photometric Calibration and Stellar
  Classification}.
\newblock {\em \aj}, {\bf 142}(Oct.), 112.

\bibitem[\protect\citename{{Buzasi} {et~al.}, }2000]{2000ApJ...532L.133B}
{Buzasi}, D., {Catanzarite}, J., {Laher}, R., {et~al.} 2000.
\newblock {The Detection of Multimodal Oscillations on {$\alpha$} Ursae
  Majoris}.
\newblock {\em \apjl}, {\bf 532}(Apr.), L133--L136.

\bibitem[\protect\citename{{Campante} {et~al.}, }2011]{2011A&A...534A...6C}
{Campante}, T.~L., {Handberg}, R., {Mathur}, S., {et~al.} 2011.
\newblock {Asteroseismology from multi-month Kepler photometry: the evolved
  Sun-like stars KIC 10273246 and KIC 10920273}.
\newblock {\em \aap}, {\bf 534}(Oct.), A6.

\bibitem[\protect\citename{{Carrier} {et~al.}, }2005]{2005A&A...434.1085C}
{Carrier}, F., {Eggenberger}, P., and {Bouchy}, F. 2005.
\newblock {New seismological results on the G0 IV {$\eta$} Bootis}.
\newblock {\em \aap}, {\bf 434}(May), 1085--1095.

\bibitem[\protect\citename{{Ceillier} {et~al.}, }2013]{2013A&A...555A..54C}
{Ceillier}, T., {Eggenberger}, P., {Garc{\'{\i}}a}, R.~A., and {Mathis}, S.
  2013.
\newblock {Understanding angular momentum transport in red giants: the case of
  KIC 7341231}.
\newblock {\em \aap}, {\bf 555}(July), A54.

\bibitem[\protect\citename{{Chaplin} {et~al.}, }2011]{2011Sci...332..213C}
{Chaplin}, W.~J., {Kjeldsen}, H., {Christensen-Dalsgaard}, J., {et~al.} 2011.
\newblock {Ensemble Asteroseismology of Solar-Type Stars with the NASA Kepler
  Mission}.
\newblock {\em Science}, {\bf 332}(Apr.), 213--.

\bibitem[\protect\citename{{Christensen-Dalsgaard}, }2004]{2004SoPh..220..137C}
{Christensen-Dalsgaard}, J. 2004.
\newblock {Physics of solar-like oscillations}.
\newblock {\em \solphys}, {\bf 220}(Apr.), 137--168.

\bibitem[\protect\citename{{Corsaro} {et~al.}, }2012]{2012ApJ...757..190C}
{Corsaro}, E., {Stello}, D., {Huber}, D., {et~al.} 2012.
\newblock {Asteroseismology of the Open Clusters NGC 6791, NGC 6811, and NGC
  6819 from 19 Months of Kepler Photometry}.
\newblock {\em \apj}, {\bf 757}(Oct.), 190.

\bibitem[\protect\citename{{De Ridder} {et~al.}, }2006]{2006A&A...448..689D}
{De Ridder}, J., {Barban}, C., {Carrier}, F., {et~al.} 2006.
\newblock {Discovery of solar-like oscillations in the red giant
  $\backslash$varepsilon Ophiuchi}.
\newblock {\em \aap}, {\bf 448}(Mar.), 689--695.

\bibitem[\protect\citename{{De Ridder} {et~al.}, }2009]{2009Natur.459..398D}
{De Ridder}, J., {Barban}, C., {Baudin}, F., {et~al.} 2009.
\newblock {Non-radial oscillation modes with long lifetimes in giant stars}.
\newblock {\em \nat}, {\bf 459}(May), 398--400.

\bibitem[\protect\citename{{Deheuvels} {et~al.}, }2010]{2010A&A...515A..87D}
{Deheuvels}, S., {Bruntt}, H., {Michel}, E., {et~al.} 2010.
\newblock {Seismic and spectroscopic characterization of the solar-like
  pulsating CoRoT target HD 49385}.
\newblock {\em \aap}, {\bf 515}(June), A87.

\bibitem[\protect\citename{{Deheuvels} {et~al.}, }2012]{2012ApJ...756...19D}
{Deheuvels}, S., {Garc{\'{\i}}a}, R.~A., {Chaplin}, W.~J., {et~al.} 2012.
\newblock {Seismic Evidence for a Rapidly Rotating Core in a Lower-giant-branch
  Star Observed with Kepler}.
\newblock {\em \apj}, {\bf 756}(Sept.), 19.

\bibitem[\protect\citename{{Deheuvels} {et~al.}, }2014]{2014arXiv1401.3096D}
{Deheuvels}, S., {Do{\u g}an}, G., {Goupil}, M.~J., {et~al.} 2014.
\newblock {Seismic constraints on the radial dependence of the internal
  rotation profiles of six Kepler subgiants and young red giants}.
\newblock {\em ArXiv e-prints}, Jan.

\bibitem[\protect\citename{{Dupret} {et~al.}, }2009]{2009A&A...506...57D}
{Dupret}, {M.-A.}, {Belkacem}, K., {Samadi}, R., {et~al.} 2009.
\newblock {Theoretical amplitudes and lifetimes of non-radial solar-like
  oscillations in red giants}.
\newblock {\em \aap}, {\bf 506}(Oct.), 57--67.

\bibitem[\protect\citename{{Dziembowski} and {Pamyatnykh},
  }1991]{1991A&A...248L..11D}
{Dziembowski}, W.~A., and {Pamyatnykh}, A.~A. 1991.
\newblock {A potential asteroseismological test for convective overshooting
  theories}.
\newblock {\em \aap}, {\bf 248}(Aug.), L11--L14.

\bibitem[\protect\citename{{Eggenberger} {et~al.}, }2012]{2012A&A...544L...4E}
{Eggenberger}, P., {Montalb{\'a}n}, J., and {Miglio}, A. 2012.
\newblock {Angular momentum transport in stellar interiors constrained by
  rotational splittings of mixed modes in red giants}.
\newblock {\em \aap}, {\bf 544}(Aug.), L4.

\bibitem[\protect\citename{{Frandsen} {et~al.}, }2002]{2002A&A...394L...5F}
{Frandsen}, S., {Carrier}, F., {Aerts}, C., {et~al.} 2002.
\newblock {Detection of Solar-like oscillations in the G7 giant star xi Hya}.
\newblock {\em \aap}, {\bf 394}(Oct.), L5--L8.

\bibitem[\protect\citename{{Garc{\'{\i}}a} {et~al.},
  }2007]{2007Sci...316.1591G}
{Garc{\'{\i}}a}, R.~A., {Turck-Chi\`eze}, S., {Jim\'enez-Reyes}, S.~J.,
  {et~al.} 2007.
\newblock {Tracking Solar Gravity Modes: The Dynamics of the Solar Core}.
\newblock {\em Science}, {\bf 316}(June), 1591--1593.

\bibitem[\protect\citename{{Garc{\'{\i}}a} {et~al.},
  }2014]{2014A&A...563A..84G}
{Garc{\'{\i}}a}, R.~A., {P{\'e}rez Hern{\'a}ndez}, F., {Benomar}, O., {et~al.}
  2014.
\newblock {Study of KIC 8561221 observed by Kepler: an early red giant showing
  depressed dipolar modes}.
\newblock {\em \aap}, {\bf 563}(Mar.), A84.

\bibitem[\protect\citename{{Gilliland}, }2008]{2008AJ....136..566G}
{Gilliland}, R.~L. 2008.
\newblock {Photometric Oscillations of Low-Luminosity Red Giant Stars}.
\newblock {\em \aj}, {\bf 136}(Aug.), 566--579.

\bibitem[\protect\citename{{Gilliland} {et~al.}, }1993]{1993AJ....106.2441G}
{Gilliland}, R.~L., {Brown}, T.~M., {Kjeldsen}, H., {et~al.} 1993.
\newblock {A search for solar-like oscillations in the stars of M67 with CCD
  ensemble photometry on a network of 4 M telescopes}.
\newblock {\em \aj}, {\bf 106}(Dec.), 2441--2476.

\bibitem[\protect\citename{{Gilliland} {et~al.}, }2010]{2010PASP..122..131G}
{Gilliland}, R.~L., {Brown}, T.~M., {Christensen-Dalsgaard}, J., {et~al.} 2010.
\newblock {Kepler Asteroseismology Program: Introduction and First Results}.
\newblock {\em \pasp}, {\bf 122}(Feb.), 131--143.

\bibitem[\protect\citename{{Goldreich} and {Keeley},
  }1977]{1977ApJ...212..243G}
{Goldreich}, P., and {Keeley}, D.~A. 1977.
\newblock {Solar seismology. II - The stochastic excitation of the solar
  p-modes by turbulent convection}.
\newblock {\em \apj}, {\bf 212}(Feb.), 243--251.

\bibitem[\protect\citename{{Goupil} {et~al.}, }2013]{2013A&A...549A..75G}
{Goupil}, M.~J., {Mosser}, B., {Marques}, J.~P., {et~al.} 2013.
\newblock {Seismic diagnostics for transport of angular momentum in stars. II.
  Interpreting observed rotational splittings of slowly rotating red giant
  stars}.
\newblock {\em \aap}, {\bf 549}(Jan.), A75.

\bibitem[\protect\citename{{Hekker} {et~al.}, }2011]{2011MNRAS.414.2594H}
{Hekker}, S., {Gilliland}, R.~L., {Elsworth}, Y., {et~al.} 2011.
\newblock {Characterization of red giant stars in the public Kepler data}.
\newblock {\em \mnras}, {\bf 414}(July), 2594--2601.

\bibitem[\protect\citename{{Huber} {et~al.}, }2011]{2011ApJ...743..143H}
{Huber}, D., {Bedding}, T.~R., {Stello}, D., {et~al.} 2011.
\newblock {Testing Scaling Relations for Solar-like Oscillations from the Main
  Sequence to Red Giants Using Kepler Data}.
\newblock {\em \apj}, {\bf 743}(Dec.), 143.

\bibitem[\protect\citename{{Huber} {et~al.}, }2014]{2014ApJS..211....2H}
{Huber}, D., {Silva Aguirre}, V., {Matthews}, J.~M., {et~al.} 2014.
\newblock {Revised Stellar Properties of Kepler Targets for the Quarter 1-16
  Transit Detection Run}.
\newblock {\em \apjs}, {\bf 211}(Mar.), 2.

\bibitem[\protect\citename{{Iben}, }1971]{1971PASP...83..697I}
{Iben}, Jr., I. 1971.
\newblock {Globular Cluster Stars: Results of Theoretical Evolution and
  Pulsation Studies Compared with the Observations}.
\newblock {\em \pasp}, {\bf 83}(Dec.), 697.

\bibitem[\protect\citename{{Kallinger} {et~al.}, }2010]{2010A&A...522A...1K}
{Kallinger}, T., {Mosser}, B., {Hekker}, S., {et~al.} 2010.
\newblock {Asteroseismology of red giants from the first four months of Kepler
  data: Fundamental stellar parameters}.
\newblock {\em \aap}, {\bf 522}(Nov.), A1.

\bibitem[\protect\citename{{Kallinger} {et~al.}, }2012]{2012A&A...541A..51K}
{Kallinger}, T., {Hekker}, S., {Mosser}, B., {et~al.} 2012.
\newblock {Evolutionary influences on the structure of red-giant acoustic
  oscillation spectra from 600d of Kepler observations}.
\newblock {\em \aap}, {\bf 541}(May), A51.

\bibitem[\protect\citename{{Kawaler} {et~al.}, }1999]{1999ApJ...516..349K}
{Kawaler}, S.~D., {Sekii}, T., and {Gough}, D. 1999.
\newblock {Prospects for Measuring Differential Rotation in White Dwarfs
  through Asteroseismology}.
\newblock {\em \apj}, {\bf 516}(May), 349--365.

\bibitem[\protect\citename{{Kjeldsen} and {Bedding},
  }1995]{1995A&A...293...87K}
{Kjeldsen}, H., and {Bedding}, T.~R. 1995.
\newblock {Amplitudes of stellar oscillations: the implications for
  asteroseismology.}
\newblock {\em \aap}, {\bf 293}(Jan.), 87--106.

\bibitem[\protect\citename{{Kjeldsen} {et~al.}, }1995]{1995AJ....109.1313K}
{Kjeldsen}, H., {Bedding}, T.~R., {Viskum}, M., and {Frandsen}, S. 1995.
\newblock {Solarlike oscillations in eta Boo}.
\newblock {\em \aj}, {\bf 109}(Mar.), 1313--1319.

\bibitem[\protect\citename{{Leighton} {et~al.}, }1962]{1962ApJ...135..474L}
{Leighton}, R.~B., {Noyes}, R.~W., and {Simon}, G.~W. 1962.
\newblock {Velocity Fields in the Solar Atmosphere. I. Preliminary Report.}
\newblock {\em \apj}, {\bf 135}(Mar.), 474.

\bibitem[\protect\citename{{Marques} {et~al.}, }2013]{2013A&A...549A..74M}
{Marques}, J.~P., {Goupil}, M.~J., {Lebreton}, Y., {et~al.} 2013.
\newblock {Seismic diagnostics for transport of angular momentum in stars. I.
  Rotational splittings from the pre-main sequence to the red-giant branch}.
\newblock {\em \aap}, {\bf 549}(Jan.), A74.

\bibitem[\protect\citename{{Mathur} {et~al.}, }2011]{2011ApJ...733...95M}
{Mathur}, S., {Handberg}, R., {Campante}, T.~L., {et~al.} 2011.
\newblock {Solar-like Oscillations in KIC 11395018 and KIC 11234888 from 8
  Months of Kepler Data}.
\newblock {\em \apj}, {\bf 733}(June), 95.

\bibitem[\protect\citename{{Mathur} {et~al.}, }2012]{2012ApJ...749..152M}
{Mathur}, S., {Metcalfe}, T.~S., {Woitaszek}, M., {et~al.} 2012.
\newblock {A Uniform Asteroseismic Analysis of 22 Solar-type Stars Observed by
  Kepler}.
\newblock {\em \apj}, {\bf 749}(Apr.), 152.

\bibitem[\protect\citename{{Metcalfe} {et~al.}, }2010]{2010ApJ...723.1583M}
{Metcalfe}, T.~S., {Monteiro}, M.~J.~P.~F.~G., {Thompson}, M.~J., {et~al.}
  2010.
\newblock {A Precise Asteroseismic Age and Radius for the Evolved Sun-like Star
  KIC 11026764}.
\newblock {\em \apj}, {\bf 723}(Nov.), 1583--1598.

\bibitem[\protect\citename{{Miglio} {et~al.}, }2009]{2009A&A...503L..21M}
{Miglio}, A., {Montalb{\'a}n}, J., {Baudin}, F., {et~al.} 2009.
\newblock {Probing populations of red giants in the galactic disk with CoRoT}.
\newblock {\em \aap}, {\bf 503}(Sept.), L21--L24.

\bibitem[\protect\citename{{Miglio} {et~al.}, }2012a]{2012MNRAS.419.2077M}
{Miglio}, A., {Brogaard}, K., {Stello}, D., {et~al.} 2012a.
\newblock {Asteroseismology of old open clusters with Kepler: direct estimate
  of the integrated red giant branch mass-loss in NGC 6791 and 6819}.
\newblock {\em \mnras}, {\bf 419}(Jan.), 2077--2088.

\bibitem[\protect\citename{{Miglio} {et~al.}, }2012b]{2012EPJWC..1905012M}
{Miglio}, A., {Morel}, T., {Barbieri}, M., {et~al.} 2012b.
\newblock {Solar-like pulsating stars as distance indicators: G-K giants in the
  CoRoT and Kepler fields}.
\newblock {\em Assembling the Puzzle of the Milky Way, Le Grand-Bornand,
  France, Edited by C.~Reyl{\'e}; A.~Robin; M.~Schultheis; EPJ Web of
  Conferences, Volume 19, id.05012}, {\bf 19}(Feb.), 5012.

\bibitem[\protect\citename{{Miglio} {et~al.}, }2013a]{2013EPJWC..4303004M}
{Miglio}, A., {Chiappini}, C., {Morel}, T., {et~al.} 2013a (Mar.).
\newblock {Differential population studies using asteroseismology: Solar-like
  oscillating giants in CoRoT fields LRc01 and LRa01}.
\newblock {Page  3004 of:} {\em European Physical Journal Web of Conferences}.
\newblock European Physical Journal Web of Conferences, vol. 43.

\bibitem[\protect\citename{{Miglio} {et~al.}, }2013b]{2013MNRAS.429..423M}
{Miglio}, A., {Chiappini}, C., {Morel}, T., {et~al.} 2013b.
\newblock {Galactic archaeology: mapping and dating stellar populations with
  asteroseismology of red-giant stars}.
\newblock {\em \mnras}, {\bf 429}(Feb.), 423--428.

\bibitem[\protect\citename{{Montalb{\'a}n} {et~al.},
  }2013]{2013ApJ...766..118M}
{Montalb{\'a}n}, J., {Miglio}, A., {Noels}, A., {et~al.} 2013.
\newblock {Testing Convective-core Overshooting Using Period Spacings of Dipole
  Modes in Red Giants}.
\newblock {\em \apj}, {\bf 766}(Apr.), 118.

\bibitem[\protect\citename{{Mosser} {et~al.}, }2010]{2010A&A...517A..22M}
{Mosser}, B., {Belkacem}, K., {Goupil}, {M.-J.}, {et~al.} 2010.
\newblock {Red-giant seismic properties analyzed with CoRoT}.
\newblock {\em \aap}, {\bf 517}(July), A22.

\bibitem[\protect\citename{{Mosser} {et~al.}, }2011]{2011A&A...532A..86M}
{Mosser}, B., {Barban}, C., {Montalb{\'a}n}, J., {et~al.} 2011.
\newblock {Mixed modes in red-giant stars observed with CoRoT}.
\newblock {\em \aap}, {\bf 532}(Aug.), A86.

\bibitem[\protect\citename{{Mosser} {et~al.}, }2012a]{2012A&A...537A..30M}
{Mosser}, B., {Elsworth}, Y., {Hekker}, S., {et~al.} 2012a.
\newblock {Characterization of the power excess of solar-like oscillations in
  red giants with Kepler}.
\newblock {\em \aap}, {\bf 537}(Jan.), A30.

\bibitem[\protect\citename{{Mosser} {et~al.}, }2012b]{2012A&A...548A..10M}
{Mosser}, B., {Goupil}, M.~J., {Belkacem}, K., {et~al.} 2012b.
\newblock {Spin down of the core rotation in red giants}.
\newblock {\em \aap}, {\bf 548}(Dec.), A10.

\bibitem[\protect\citename{{Mosser} {et~al.}, }2013]{2013A&A...559A.137M}
{Mosser}, B., {Dziembowski}, W.~A., {Belkacem}, K., {et~al.} 2013.
\newblock {Period-luminosity relations in evolved red giants explained by
  solar-like oscillations}.
\newblock {\em \aap}, {\bf 559}(Nov.), A137.

\bibitem[\protect\citename{{Osaki}, }1975]{1975PASJ...27..237O}
{Osaki}, J. 1975.
\newblock {Nonradial oscillations of a 10 solar mass star in the main-sequence
  stage}.
\newblock {\em \pasj}, {\bf 27}, 237--258.

\bibitem[\protect\citename{{Paxton} {et~al.}, }2011]{2011ApJS..192....3P}
{Paxton}, B., {Bildsten}, L., {Dotter}, A., {et~al.} 2011.
\newblock {Modules for Experiments in Stellar Astrophysics (MESA)}.
\newblock {\em \apjs}, {\bf 192}(Jan.), 3.

\bibitem[\protect\citename{{Paxton} {et~al.}, }2013]{2013ApJS..208....4P}
{Paxton}, B., {Cantiello}, M., {Arras}, P., {et~al.} 2013.
\newblock {Modules for Experiments in Stellar Astrophysics (MESA): Planets,
  Oscillations, Rotation, and Massive Stars}.
\newblock {\em \apjs}, {\bf 208}(Sept.), 4.

\bibitem[\protect\citename{{Rauer} {et~al.}, }2013]{2013arXiv1310.0696R}
{Rauer}, H., {Catala}, C., {Aerts}, C., {et~al.} 2013.
\newblock {The PLATO 2.0 Mission}.
\newblock {\em ArXiv e-prints}, Oct.

\bibitem[\protect\citename{{Retter} {et~al.}, }2003]{2003ApJ...591L.151R}
{Retter}, A., {Bedding}, T.~R., {Buzasi}, D.~L., {Kjeldsen}, H., and {Kiss},
  L.~L. 2003.
\newblock {Oscillations in Arcturus from WIRE Photometry}.
\newblock {\em \apjl}, {\bf 591}(July), L151--L154.

\bibitem[\protect\citename{{Ricker} {et~al.}, }2010]{2010AAS...21545006R}
{Ricker}, G.~R., {Latham}, D.~W., {Vanderspek}, R.~K., {et~al.} 2010 (Jan.).
\newblock {Transiting Exoplanet Survey Satellite (TESS)}.
\newblock {Page  450.06 of:} {\em American Astronomical Society Meeting
  Abstracts 215}.
\newblock Bulletin of the American Astronomical Society, vol. 42.

\bibitem[\protect\citename{{Sills} and {Pinsonneault},
  }2000]{2000ApJ...540..489S}
{Sills}, A., and {Pinsonneault}, M.~H. 2000.
\newblock {Rotation of Horizontal-Branch Stars in Globular Clusters}.
\newblock {\em \apj}, {\bf 540}(Sept.), 489--503.

\bibitem[\protect\citename{{Silva Aguirre} {et~al.},
  }2012]{2012ApJ...757...99S}
{Silva Aguirre}, V., {Casagrande}, L., {Basu}, S., {et~al.} 2012.
\newblock {Verifying Asteroseismically Determined Parameters of Kepler Stars
  Using Hipparcos Parallaxes: Self-consistent Stellar Properties and
  Distances}.
\newblock {\em \apj}, {\bf 757}(Sept.), 99.

\bibitem[\protect\citename{{Stello} {et~al.}, }2006]{2006A&A...448..709S}
{Stello}, D., {Kjeldsen}, H., {Bedding}, T.~R., and {Buzasi}, D. 2006.
\newblock {Oscillation mode lifetimes in {$\xi$} Hydrae: will strong mode
  damping limit asteroseismology of red giant stars?}
\newblock {\em \aap}, {\bf 448}(Mar.), 709--715.

\bibitem[\protect\citename{{Stello} {et~al.}, }2007]{2007MNRAS.377..584S}
{Stello}, D., {Bruntt}, H., {Kjeldsen}, H., {et~al.} 2007.
\newblock {Multisite campaign on the open cluster M67 - II. Evidence for
  solar-like oscillations in red giant stars}.
\newblock {\em \mnras}, {\bf 377}(May), 584--594.

\bibitem[\protect\citename{{Stello} {et~al.}, }2008]{2008ApJ...674L..53S}
{Stello}, D., {Bruntt}, H., {Preston}, H., and {Buzasi}, D. 2008.
\newblock {Oscillating K Giants with the WIRE Satellite: Determination of Their
  Asteroseismic Masses}.
\newblock {\em \apjl}, {\bf 674}(Feb.), L53--L56.

\bibitem[\protect\citename{{Stello} {et~al.}, }2009]{2009MNRAS.400L..80S}
{Stello}, D., {Chaplin}, W.~J., {Basu}, S., {Elsworth}, Y., and {Bedding},
  T.~R. 2009.
\newblock {The relation between {$\Delta$}{$\nu$} and {$\nu$}$_{max}$ for
  solar-like oscillations}.
\newblock {\em \mnras}, {\bf 400}(Nov.), L80--L84.

\bibitem[\protect\citename{{Stello} {et~al.}, }2011]{2011ApJ...739...13S}
{Stello}, D., {Meibom}, S., {Gilliland}, R.~L., {et~al.} 2011.
\newblock {An Asteroseismic Membership Study of the Red Giants in Three Open
  Clusters Observed by Kepler: NGC 6791, NGC 6819, and NGC 6811}.
\newblock {\em \apj}, {\bf 739}(Sept.), 13.

\bibitem[\protect\citename{{Stello} {et~al.}, }2013]{2013ApJ...765L..41S}
{Stello}, D., {Huber}, D., {Bedding}, T.~R., {et~al.} 2013.
\newblock {Asteroseismic Classification of Stellar Populations among 13,000 Red
  Giants Observed by Kepler}.
\newblock {\em \apjl}, {\bf 765}(Mar.), L41.

\bibitem[\protect\citename{{van Leeuwen}, }2007]{2007ASSL..350.....V}
{van Leeuwen}, F. 2007.
\newblock {\em {Hipparcos, the New Reduction of the Raw Data}}.
\newblock Astrophysics and Space Science Library, vol. 350.
\newblock { }.

\bibitem[\protect\citename{{White} {et~al.}, }2011]{2011ApJ...743..161W}
{White}, T.~R., {Bedding}, T.~R., {Stello}, D., {et~al.} 2011.
\newblock {Calculating Asteroseismic Diagrams for Solar-like Oscillations}.
\newblock {\em \apj}, {\bf 743}(Dec.), 161.

\bibitem[\protect\citename{{White} {et~al.}, }2013]{2013MNRAS.433.1262W}
{White}, T.~R., {Huber}, D., {Maestro}, V., {et~al.} 2013.
\newblock {Interferometric radii of bright Kepler stars with the CHARA Array:
  theta Cygni and 16 Cygni A and B}.
\newblock {\em \mnras}, {\bf 433}(Aug.), 1262--1270.

\end{thebibliography}

\end{document}